\documentclass[ reprint, showpacs, amsmath, amssymb, aps, pra, floatfix,]{revtex4-1}
\pdfoutput=1

\usepackage{setspace}
\usepackage{enumerate}
\usepackage{graphicx}
\graphicspath{ {./figures/} }

\usepackage{dcolumn}
\usepackage{bm}

\def\Res{\mathop{\mathrm{Res}}}
\providecommand{\bra}[1]{\left\langle {#1}\right\vert}
\providecommand{\ket}[1]{\left\vert {#1}\right\rangle}
\providecommand{\mat}[1]{\mathcal{#1}}

\providecommand{\EXP}[1]{\mathrm{e}^{#1}}
\providecommand{\abs}[1]{\left\vert{#1}\right\vert}
\providecommand{\I}{\mathrm{i}}
\providecommand{\D}{\mathrm{d}}
\providecommand{\myspacer}{\rule{14pt}{0pt}}

\begin{document}

\title{Weak Limit of the 3-State Quantum Walk on the Line}
\author{Stefan Falkner}
	\affiliation{%
	Department  of Physics, Emory University, Atlanta, GA, 30322; USA%
	}
\author{Stefan Boettcher}
	\affiliation{%
	Department  of Physics, Emory University, Atlanta, GA, 30322; USA%
	}

\begin{abstract}
We revisit the one dimensional discrete time quantum walk with 3-states and the Grover coin, the simplest model that exhibits localization in a quantum walk. We derive analytic expressions for the localization and a long-time approximation for the entire probability density function (PDF). We also connect the time-averaged approximation of the PDF found by Inui et.~al.~to a spatial average of the walk. We show that this smoothed approximation predicts moments of the real PDF accurately. 
\end{abstract}


\maketitle

\begin{figure}[b!]
\vspace{-.3in}
\centering
\includegraphics{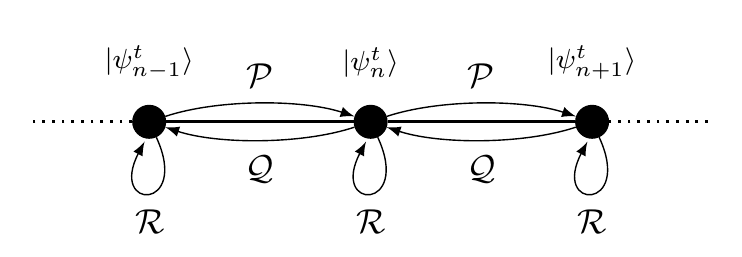}
\caption{The one dimensional quantum walk with a three dimensional coin. The matrices $\mat P$, $\mat Q$ and $\mat R$ facilitate a right hop, a left hop, or no change in position.}
\label{fig:lattice}
\end{figure}
\section{Introduction}
Quantum walks have been the subject of research for the past 20 years~\cite{Kempe03,Konno08,VA12,PortugalBook}. They were originally proposed as a description for quantum transport in a one dimensional system~\cite{aharonov_1993a,Meyer96}. However, quantum walks soon received considerable prominence as the driving dynamics behind quantum search algorithms~\cite{Gro97a,PhysRevA.58.915}, leading to many systematic studies of their asymptotic properties~\cite{aharonov_2001a,ambainis_2001a,bach_2004a,childs_2004a}. Other applications, such as to the graph isomorphism problem~\cite{ambainis_2003a,Rudinger13}, further increased the interest. The realization of their capability for general quantum computations~\cite{childs_2009a, lovett_2010a} suggests that understanding quantum walks is a quest for a better understanding of quantum computing itself.

Due to its wealth of possible parameters, the discrete-time quantum walk has been studied extensively.  From the basic properties of the simplest possible quantum walk on the one dimensional line~\cite{ambainis_2001a},  time-dependent coins~\cite{banuls_2006a} and site-dependent coins~\cite{shikano_2010a} are just some of the many extensions that have been investigated. In this paper, we revisit the one-dimensional quantum walk with the three-dimensional Grover coin, previously considered by Inui et.~al.~\cite{inui_2005a}. They discussed a variation of the walk on the line, where the walker can remain on the site during a time step, and found interesting differences to the case of a two-dimensional coin. Most notably, there is a finite probability that the quantum walk strongly localizes around the initial site, as previously found on square lattices~\cite{inui_2004a}. In fact, this model is the simplest model exhibiting localization, a distinctly quantum effect entirely absent in the corresponding classical random walks, that becomes a generic features of discrete-time quantum walks on higher-dimensional structures~\cite{PortugalBook,QWNComms13,Yusuke14}. Here, we extend the findings of Ref.~\cite{inui_2005a} by analytic expressions for this localization, calculate the weak limit of the probability density function (PDF), and show its equivalence to a spatial average over a local neighborhood. We provide explicit expressions for general initial conditions present on one site, study the convergence, and compare our analytic predictions for moments with those from numerical simulations.

The paper is organized as follows. In Sec.~\ref{sec:3_state_qw}, we review the basics for the 3-state quantum walk on the line. In Sec.~\ref{sec:long_t_approximation} we show how the long-time behavior can be obtained, with an accurate description of the localization and an approximation for the spreading front. In Sec.~\ref{sec:weak_limit_pdf}, we introduce an approximation that leads to a smoothed PDF,  corresponding to a spatial as well as temporal average. Finally, in Sec.~\ref{sec:conclusion}, we summarize our findings.

\section{The 3-State Quantum Walk}
\label{sec:3_state_qw}

In the common description of the discrete-time quantum walk, every time step consists of two parts. First, the coin (operator) is applied to the internal degree of freedom (coin state) at every site. This is followed by the shift operator, translating components of the coin state to neighboring sites. Here, we study the case of the one dimensional quantum walk with a three dimensional coin space, driven by the Grover Coin
\begin{equation}
\label{eq:grover_coin}
\mat C = \frac{1}{3}
\begin{bmatrix} -1 & 2 & 2 \\ 2&-1&2 \\ 2&2&-1 \end{bmatrix}\,.
\end{equation}
Our convention for the shift operation is the following: the first component is moved to the left, the third component to the right, while the second one remains on the site.  The matrices $\mat P$, $\mat Q$, and $\mat R$ (see Fig.~\ref{fig:lattice}) combine both steps into a single operation, leading to the master equation describing the time evolution at any site $n$, 
\begin{equation}
\ket{\psi_n^{t+1}}= \mat P \ket{\psi_{n-1}^t} + \mat Q \ket{\psi_{n+1}^t} + \mat R \ket{\psi_n^t},
\label{eq:master_equation}
\end{equation}
with
\[
\mat P = \begin{bmatrix}0&0&0\\ 0&0&0\\ \frac23 & \frac23 & -\frac13\end{bmatrix},\,
\mat Q = \begin{bmatrix}-\frac13&\frac23&\frac23\\ 0&0&0\\ 0&0&0\end{bmatrix}\,,
\mat R = \begin{bmatrix}0&0&0\\ \frac23&-\frac13&\frac23\\ 0&0&0\end{bmatrix}\,.
\]
For simplicity, we assume the inital conditions are only non-zero on site $n=0$, i.e.\,,
\begin{equation}
\ket{\psi_n^0} = \delta_{n,0} \cdot \ket{\psi_0^0}\,.
\label{eq:initial_condition}
\end{equation}

\begin{figure}[b!]
\vspace{-.3in}
\flushright
\includegraphics{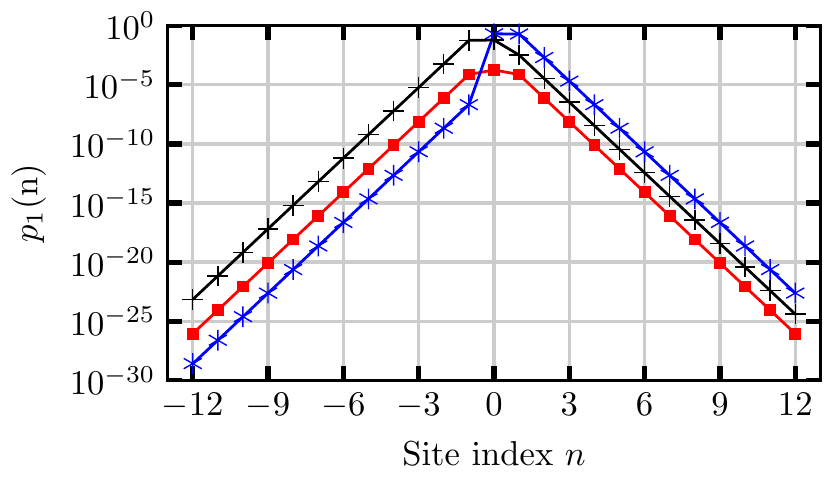}\myspacer
\caption{Comparison between analytic prediction (lines) and numerical simulation after $T=2^{20}$ time steps (symbols) for the localization around $n=0$ . The initial conditions are $\ket{\psi_0^0} \propto (1,-1.9, 1)$ (red squares),  $\ket{\psi_0^0} \propto (10,0, 1)$ (blue asterisks), and $\ket{\psi_0^0} \propto (1,-3,2+\I)$ (black pluses). They have been chosen to show the possible asymmetry of $p_1(n)$.}
\label{fig:localization_comparison}
\end{figure}

These equations can be solved by a Fourier transform,
\begin{equation}
\ket{\tilde{\psi}^t} = \sum_{n=-\infty}^\infty \EXP{-\I\cdot k\cdot n}\cdot  \ket{\psi^t_n}\,.
\label{eq:def_fourier_transform}
\end{equation}
From here on, a tilde indicates quantities with a $k$-dependence, which we will not explicitly show.  Applying Eq.~(\ref{eq:def_fourier_transform}) to Eq.~(\ref{eq:master_equation}) yields the master equation in Fourier space:
\begin{equation}
\ket{\tilde{\psi}^{t+1}}= \underbrace{\frac{1}{3}
\begin{bmatrix}
-\kappa &2\kappa & 2\kappa\\
2& -1 & 2\\
2\kappa^{-1}& 2\kappa^{-1} & -\kappa^{-1}
\end{bmatrix}}_{:=\tilde{C}} \cdot 
\ket{\tilde{\psi}^{t}}\,,
\label{eq:master_equation_in_fourier_space}
\end{equation}
where $\kappa = \EXP{\I\cdot k}$. The solution to this equation
\begin{equation}
\ket{\tilde{\psi}^{t}}= \tilde{\mat C}^t \cdot \ket{\psi_0^0}
\label{eq:solution_fourier_space}
\end{equation}
can be found by computing the eigenvalue decomposition
\begin{equation}
\mat T^{-1} \cdot \tilde{\mat C}\cdot \mat T = \begin{bmatrix} \tilde\lambda_1&0&0\\ 0&\tilde\lambda_2&0 \\ 0&0&\tilde\lambda_3
\end{bmatrix}\,.
\label{eq:C_diagonal}
\end{equation}
One eigenvalue  is purely real, $\tilde{\lambda}_1=1$, whereas the other two obey
\begin{equation}
\lambda_{2,3} = \EXP{\pm \I \tilde\omega}\quad\text{, and}\quad \cos\left( \tilde{\omega} \right) = - \frac{2}{3} - \frac{\cos(k)}{3}\,.
\label{eq:def_lambda_23}
\end{equation}
The $t^{\text{th}}$ power of $\tilde{\mat C}$ can then be expressed as
\begin{equation}
\tilde{\mat C}^t = \tilde{\mat M_1} + \tilde\lambda_2^t \cdot \tilde{\mat M_2} +  \tilde\lambda_3^t \cdot \tilde{\mat M_3}\,.
\label{eq:c_tilde_t}
\end{equation}
A representation of $\mat T$ and the $\tilde{\mat M}$ matrices can be found in the supplementary Mathematica notebook \cite{supplementary_material}. In the end, the real space solution is obtained by performing the inverse Fourier transform
\begin{equation}
\ket{\psi_n^t} = \frac{1}{2\pi} \intop_{-\pi}^{\pi} \EXP{\I\cdot n \cdot k} \cdot \ket{\tilde{\psi}^t}\, \D k\,.
\label{eq:inverse_fourier_transform}
\end{equation}
In the next section, we perform an asympotic approximation in the long-time limit to find the leading behaviour of the PDF. 

\section{Long-Time Approximation}
\label{sec:long_t_approximation}
In this section, we evaluate Eq.~(\ref{eq:inverse_fourier_transform}) in the limit of $t\to\infty$. First, we compute the time independent part of $\ket{\psi_n^t}$ that manifests itself as localization.  As a test, we compare our result with numerical simulations. Afterwards, we use the method of stationary phase to find an approximation for the remaining, time dependent part.

\subsection{The Stationary Distribution}
One can see from Eq.~(\ref{eq:c_tilde_t}) that a time-independent component of $\ket{\tilde\psi^t}$ can exist due to the constant eigenvalue of $\tilde{\mat C}$. The inverse Fourier transform of this part can be computed exactly by employing the residue theorem.  Note that the corresponding integral for this part following from equations (\ref{eq:c_tilde_t}) and (\ref{eq:inverse_fourier_transform}), in terms of $\kappa$ reads
\begin{equation}
\ket{\psi_n^\infty} = \underbrace{
		\frac{1}{2\pi\I} \ointop_{\abs\kappa=1} \kappa^{n-1} \mat{\tilde{M}}_1\,\D \kappa
	}_{:=\,\mat U_1(n)}\cdot  \ket{\psi_0^0}\,.
\label{eq:stationary_inverse_fourier}
\end{equation}
The details of this calculation can be found in Appendix~\ref{sec_a:localization}, but the essential observation is that all components of $\tilde{\mat M}_1$ share the same poles,
\begin{equation}
\kappa_{\pm} = -5 \pm 2\sqrt{6},
\label{eq:def_kappa_pm}
\end{equation}
of which only $\kappa_+$ is inside the unit circle. For $n\leq 0$, there is an additional pole at $\kappa=0$. By straightforward calculations, we find an expression for $\mat U_1(n)$ for different regimes for $n$:
\begin{align}
\notag
\mat U_1(n<0) &= \frac{\kappa_-^{n}}{\sqrt{6}}
\begin{bmatrix}
 1 & -2-\sqrt{6} & -5-2 \sqrt{6} \\
 -2+\sqrt{6} & -2 & -2-\sqrt{6} \\
 -5+2 \sqrt{6} & -2+\sqrt{6} & 1 \\
\end{bmatrix}\\
\mat U_1(n=0) &= \frac{1}{\sqrt{6}}
\begin{bmatrix}
 1 & -2+\sqrt{6} & -5+2 \sqrt{6} \\
 -2+\sqrt{6} & -2+\sqrt{6} & -2+\sqrt{6} \\
 -5+2 \sqrt{6} & -2+\sqrt{6} & 1 \\
\end{bmatrix}
\label{eq:u1_final}\\
\notag
\mat U_1(n>0) &= \frac{\kappa_+^{n}}{\sqrt{6}}
\begin{bmatrix}
 1 & -2+\sqrt{6} & -5+2 \sqrt{6} \\
 -2-\sqrt{6} & -2 & -2+\sqrt{6} \\
 -5-2 \sqrt{6} & -2-\sqrt{6} & 1 \\
\end{bmatrix}
\end{align}
At first, the case distinction in the sign of $n$ seems counterintuitive, but the comparison to the numerics in Fig.~\ref{fig:localization_comparison} reveals the possibility of an asymmetric localization around the initial site.

\begin{figure}[b!]
\vspace{-.2in}
\flushright
\includegraphics{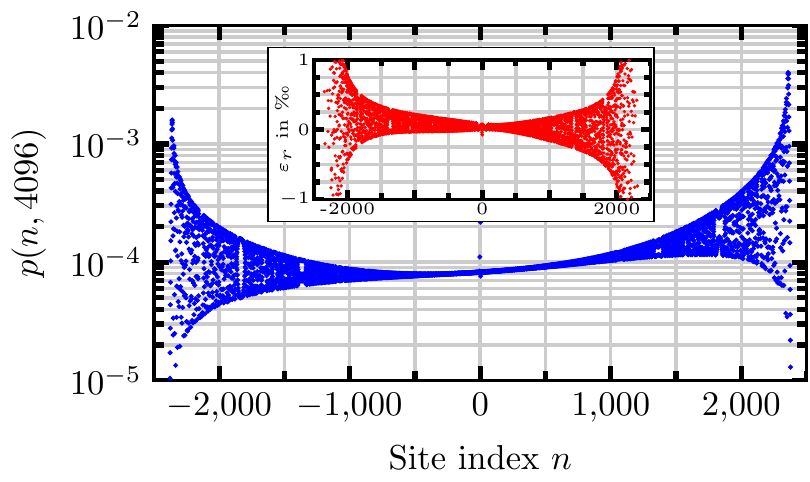}\myspacer
\caption{The PDF of the walk (blue dots) after $t=4096$ steps with the initial condition $\ket{\psi_0^0} \propto (0,\I,1)$. Beyond the shown index range, the probability is essentially zero. Inset: Relative difference between the asymptotic approximation and the numerical values. The prediction is fairly good for a wide range of points.}
\label{fig:comparison_data_pred_p_vs_n_1}
\end{figure}

To obtain the stationary PDF, we calculate
\begin{equation}
p_1(n) = \langle\psi_n^\infty)\vert\psi_n^\infty\rangle  =  \langle \psi_0^0 \vert U_1^\dagger(n) U_1(n) \vert \psi_0^0 \rangle\,.
\label{eq:def_p1}
\end{equation}
For general initial condition, $p_1(n)$ still contains the case distinction in $n$, but the localization only at the initial site for arbitrary $\ket{\psi_0^0} = ( \alpha, \beta, \gamma)^T$, for example, reads:
\begin{equation}
\begin{aligned}
&\left( \bar\alpha, \bar\beta, \bar\gamma \right) \cdot \mat U_1^\dagger(0) \cdot  \mat U_1 (0) \cdot \begin{pmatrix} \alpha\\ \beta \\ \gamma\end{pmatrix} = \left(5-2 \sqrt{6}\right)\cdot  \\ & \qquad ~~((2 \alpha +\beta ) \bar\alpha +(\alpha +\beta +\gamma ) \bar\beta+(\beta +2 \gamma ) \bar\gamma)\,,
\end{aligned}
\label{eq:localization_initial_site}
\end{equation}
which coincides with the result in Ref.~\cite{inui_2005a}.

We point out that the stationary PDF always decays exponentially away from the initial site as $p_1(n) \sim \kappa_+^{2\abs{n}}$ independent of the initial condition, even though the proportionality constant might differ from positive to negative $n$. The initial condition $\ket{\psi_0^0} \propto (1,-2,1)$ is a non-generic case where the localization  {\it completely vanishes}. There exists also a whole family,$\ket{\psi_0^0} \propto (-a\cdot (1+\kappa_\pm)/2 - b\cdot \kappa_{\pm},a,b)$ with $a,b \in \mathbb R$, where the localization vanishes for positive (negative) $n$ while still exponentially decaying for negative (positive) values.

To show that our calculations describe the localized part comprehensively, we compare to a long simulation of a system, where the system is large enough that the finite size has no influence on the PDF near the initial site at the end of the simulation. The system starts with different initial conditions and evolves for $2^{20}$ time steps. In the end, the final probabilities at sites around the origin are recorded. Figure~\ref{fig:localization_comparison} shows the comparison between evaluating Eq.~(\ref{eq:def_p1}) and the simulation. To demonstrate the asymmetry, we choose 3 particular initial conditions. 

This rapid decay renders estimating $p_1(n)$ with simulations for $\abs{n}>12$ problematic. The values range over 30 orders of magnitude, challenging the machine precision used in the simulations. Furthermore, the time to converge to $p_1(n)$ grows exponentially with $n$, as we will see, which restricts the numerical evaluation, as system size would have to grow exponentially as well.

\subsection{Approximating the Time-Dependent Integrals}
After solving the time independent part analytically, we have to resort to approximations for the time dependent part of $\ket{\tilde\psi^t}$ in the limit $t\to\infty$. In analogy to Eq.~(\ref{eq:stationary_inverse_fourier}), we define
\begin{equation}
\mat U_{2,3}(t,n) = \frac{1}{2\pi} \intop_{-\pi}^\pi \EXP{\I\cdot k \cdot n} \tilde{\mat{M}}_{2,3}\cdot\tilde\lambda_{2,3}^t\,\D k\,,
\label{eq:def_U23}
\end{equation}
such that the sum $\mat U_1 + \mat U_2 + \mat U_3 =\tilde{\mat C}^t$ expresses the full time evolution. By introducing the "velocity" $v$ via 
\begin{equation}
n = v\cdot t
\label{eq:def_v}
\end{equation}
and using Eq.~(\ref{eq:def_lambda_23}), we write the integrals as
\[
\frac{1}{2\pi} \intop_{-\pi}^\pi \tilde{f}(k) \cdot \EXP{\I t \cdot \left( v k \pm \tilde{\omega} \right)}\,\D k\,.
\]
where the function $\tilde{f}$ represents the different (slowly-varying) elements of the $\tilde{\mat M}$ matrices.

This form is known as a generalized Fourier integral~\cite{bender_1999a}, and the leading, long-time behavior can be found by the method of stationary phase. The method assumes that the main contribution to the integral stems from a small region of $k$ around an extremal value of $(vk\pm \tilde{\omega})$, say $k^*$. Expanding the exponent to second order and replacing the function $\tilde{f}(k)$ by $\tilde{f}(k^*)$ yields a solvable Gaussian integral. A more detailed discussion can be found in Appendix~\ref{sec_a:asymptotic_approximation}. 

In this approximation, $\tilde{\mat C}^t$ will contain the constant terms from $\mat U_1$ and terms proportional to $t^{-1/2}$ that further oscillate both in space and in time. In their full extend, these terms are too complex to write down here, but easily used to compute numerical values for specific initial conditions. The supplementary Mathematica notebook contains an applet that shows the approximation for interactive initial conditions \cite{supplementary_material}.

Figure~\ref{fig:comparison_data_pred_p_vs_n_1} shows the PDF for a specific initial condition as a function of the site index $n$ for a fixed time $t$. To show the quality of the approximation, we also show the relative difference
\begin{equation}
\varepsilon_r = \frac{2\cdot\abs{p_{s}(n,t) - p_{a}(n,t)}}{p_{s}(n,t) + p_{a}(n,t)}
\label{eq:def_relative_difference}
\end{equation}
between the simulation $p_{s}$ and the asymptotic expression $p_{a}$. Note that prediction and simulation are indistinguishable on this scale. The quality of the prediction remains excellent for general   (complex and asymmetric) initial conditions $\ket{\psi_0^0}$.

\begin{figure}[b!]
\vspace{-.2in}
\flushright
\includegraphics{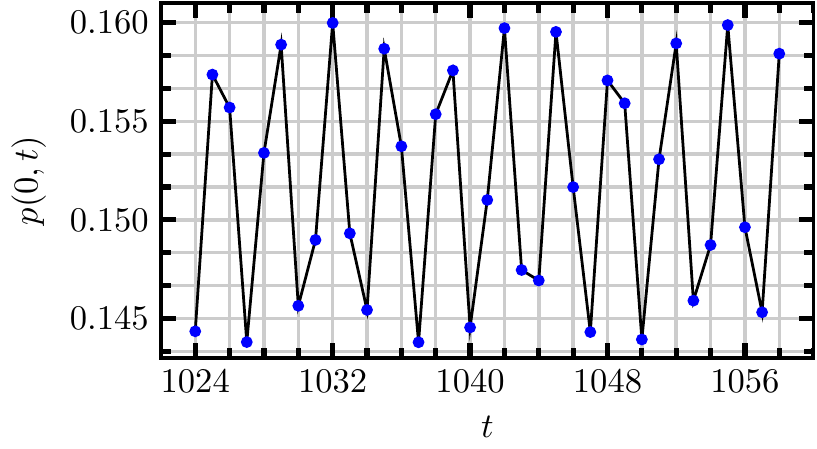}\myspacer

\includegraphics{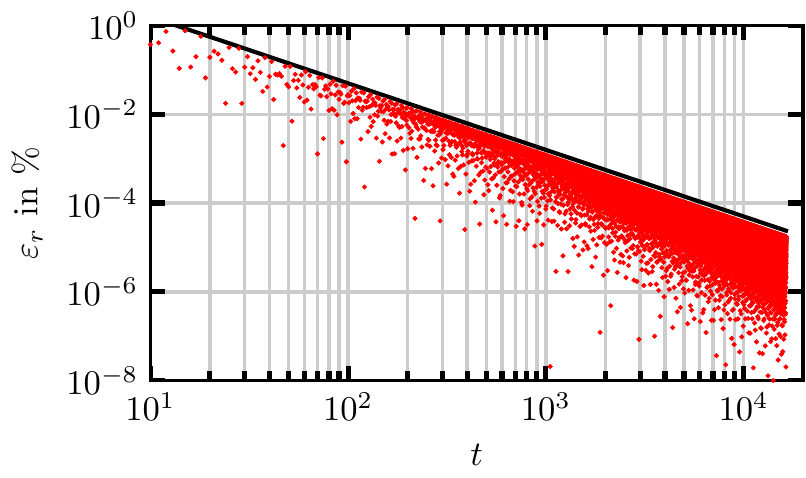}\myspacer
\caption{ {\it Top:} A short time series of the probability at the initial site. The numerical values (blue circles) and the asymptotic approximation (black line) are indistinguishable on this scale. {\it Bottom:} The relative difference for $p(0,t)$ between the simulation and asymptotic approximation. The black line corresponds to $\sim t^{-3/2}$ and is just a guide to the eye.}
\label{fig:comparison_data_pred_p_vs_t_1}
\end{figure}

The approximation can also be used for a fixed $n$ as a function of $t$, as demonstrated in Fig.~\ref{fig:comparison_data_pred_p_vs_t_1} for the initial site. The plot displays a short short sequence of a time series at large $t$. Again, simulation and asymptotic approximation are indistinguishable on this scale.

To better understand the quality of the approximation, we plot the relative difference in the bottom part of Fig.~\ref{fig:comparison_data_pred_p_vs_t_1}. The data  suggests,  surprisingly, that the error of the approximation decays as $\sim t^{-3/2}$. This would imply that the method of stationary phase correctly predicts the leading behavior to order $t^{-1}$. This cannot be expected a priory, because the next order, obtainable with the method of steepest descent, may yield terms of that order for $\tilde{\mat C}^t$.  Those should generate terms of the same magnitude in the PDF due to the constant eigenvalue. The data here suggests that such terms cancel out.

\section{Weak Limit Distribution}
\label{sec:weak_limit_pdf}
In the previous section, we found that the method of stationary phase yields a good approximation to the evolution of the quantum walk for sufficiently large times. It also became obvious that the PDF oscillates as a function of both $n$ and $t$, especially close to the moving front, near $\left|v\right|\lesssim 1/\sqrt{3}$. In this section, we find a smooth approximation known as the weak limit~\cite{grimmett_2004a}. We demonstrate that it yields a proper PDF, study the convergence towards it, and show that the walk spreads ballistically for all initial conditions.

\begin{figure}[b!]
\vspace{-.2in}
\flushright
\includegraphics{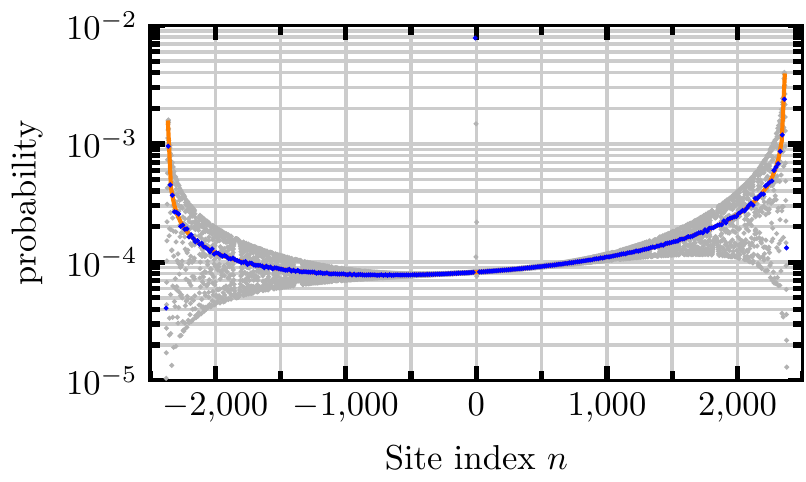}\myspacer
\caption{The smooth approximation $p_{avg}$ from Eq.~\ref{eq:definition_p_avg} (red line) and the spatial average over 16 sites (blue dots) after 4096 time steps. The initial conditions are the same as in Fig.~\ref{fig:comparison_data_pred_p_vs_n_1}. The actual PDF is shown by gray dots.}
\label{fig:avg_pdf_comparison1}
\end{figure}

\subsection{Properties and Implications}
 Following Ambainis {\it et.~al.~}\cite{ambainis_2001a}, we can separate out the rapidly oscillating part of Eq.~(\ref{eq:c_tilde_t}).  If we ignore the localized part for a moment, the corresponding distribution, which we will call $p_{avg}(n,t)$, can be found via
\begin{equation}
p_{avg}(n,t) = \bra{\psi_0^0} \left( \mat U_2^\dagger\cdot \mat U_2 + \mat U_3^\dagger\cdot \mat U_3 \right) \ket{\psi_0^0}
\label{eq:definition_p_avg}
\end{equation}
This expression seems ad hoc, but contains all non-oscillating terms from the full approximation.  This corresponds to a temporal average at a specific site, assuming that the rapidly oscillating phase factors lead to a negligible contribution to the inverse Fourier transform (according to Riemann-Lebesgue). We argue that this also corresponds to a local spatial average at fixed $t$, because as $t\to\infty$ a small change in $n$ will only lead to an small change in $v$, such that the non-oscillating contribution should be the same in a  neighborhood around a point that is reasonably small compared to $t$. This average also smoothes out the spatial oscillations of the PDF.  In fact, we will use a spatial average to numerically predict $p_{avg}$.

Inserting the expressions for $\mat U_{2,3}(n,t)$, we find the matrix
\begin{equation}
\begin{aligned}
\mat U_2^\dagger\cdot \mat U_2 + &\mat U_3^\dagger\cdot \mat U_3 =\frac{1}{\pi t \sqrt{2(1-3v^2)} ~ (1-v^2)} \cdot \\ &
\begin{bmatrix}
(1-v)^2 & 2 v (1-v) & 1-5v^2\\
2v(1-v) & 2-2v^2    & -2v(1+v)\\
1-5v^2  & -2v(1+v)  & (1+v)^2
\end{bmatrix}
\end{aligned}
\label{eq:matrix_for_p_avg}
\end{equation}
valid for all $n$ subjected to $\abs{n}/t < 1/\sqrt{3}$. Outside this interval, $p_{avg}(n,t) \equiv 0$. The dependency on $n$ is implicit through $v=n/t$.  For a specific initial condition, a comparison between the numerical simulation and the analytic prediction can be found in Fig.~\ref{fig:avg_pdf_comparison1}.

Our definition of $p_{avg}(n,t)$ closely relates to the weak limit proven by Grimmett {\it et.~al.~}\cite{grimmett_2004a}. Note that $p_{avg}(n,t)/t$ only depends on $v$ which corresponds to $f(y)$ in their notation. They show that every quantum walk on regular lattices exhibits this convergence, for example, see  Eq.~(20) in Ref.~\cite{grimmett_2004a}.

\begin{figure}[b!]
\vspace{-.3in}
\flushright
\includegraphics{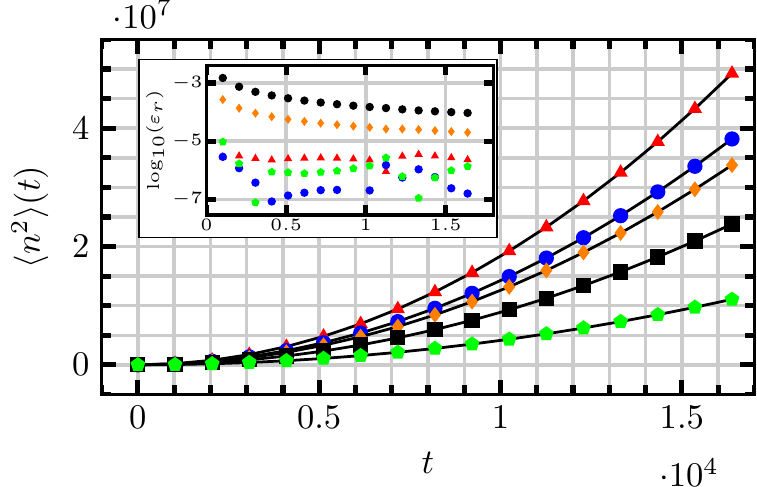} \myspacer
\caption{Comparison between the numerical values (symbols) and the evaluation of Eq.~(\ref{eq:second_moment}) for the second moment of the PDF. The inset shows the relative difference between the two values.}
\label{fig:second_moment}
\end{figure}

In the long-time limit, we can tread $v=n/t$ as a continuous variable. Hence, we approximate probabilities 
\[
p(n_-\leq n \leq n_+, t) = \sum_{n=n_-}^{n_+} p(n,t)
\]
by integrals of the form
\[
p(a\  \leq v \leq b,t) = \intop_{a}^b p_{avg}(v\cdot t, t) \cdot t\,\D v\,.
\]
Here $a=n_-/t$ and $b=n_+/t$. By using the convergence of $p_{avg}(v\cdot t,t)\cdot t$ to a stationary distribution solely depending on $v$, we conclude that the spreading is always ballistic. In this continuous limit, the localized part remains concentrated at the initial site, and
\begin{equation}
p_1(v\cdot t)\cdot t \to
\bra{\psi_0^0}
\begin{pmatrix}
 \frac{1}{\sqrt{6}} & 1-\sqrt{\frac{2}{3}} & 2-\frac{5}{\sqrt{6}} \\
 1-\sqrt{\frac{2}{3}} & 1-\sqrt{\frac{2}{3}} & 1-\sqrt{\frac{2}{3}} \\
 2-\frac{5}{\sqrt{6}} & 1-\sqrt{\frac{2}{3}} & \frac{1}{\sqrt{6}} \\
\end{pmatrix}
\ket{\psi_0^0}\, \delta(v)
\label{eq:limit_p1}
\end{equation}
characterizing the localization within the weak limit. Some algebra reveals that 
\begin{equation}
\label{eq:proper_pdf}
\sum_{n=-\infty}^\infty p_1(n) + \intop_{-1/\sqrt{3}}^{1/\sqrt{3}} p_{avg}(v\cdot t, t)\cdot t ~ \D v = 1\,,
\end{equation}
i.e., our approximation actually yields a proper PDF. Connecting once more with Ref.~\cite{inui_2004a}, if the system starts in one of the three initial states $(1,0,0)$, $(0,1,0)$, and $(0,0,1)$ each with probability $1/3$, we rediscover:
\[
p(v,t) \approx \frac{1}{3}\,\delta(v) + \frac{4}{3\pi\cdot (1-v^2) \cdot \sqrt{2-6v^2}}\,.
\]
But observe that generically the numerator of the second term is quadratic in $v$ rather than just a constant, as can be seen from Eq.~(\ref{eq:matrix_for_p_avg}).
As an example, we utilize $p_{avg}(n,t)$ to approximate the second moment of the PDF. Figure~\ref{fig:second_moment} shows a comparison between the approximation and numerical simulations. The details of the calculation are in Appendix~\ref{seq_a:moment}, but the main result is that the second moment always grows $\sim t^2$ regardless of the initial condition, see Eq.~(\ref{eq:second_moment}). This means, only the PDF's shape can be influenced by $\ket{\psi_0^0}$, but not the asymptotic scaling of its spread.

In principle, we can approximate every moment, but the quality declines for higher moments. Those depend stronger on sites farther away from the initial site where the accuracy is worse.

\begin{figure*}
\begin{tabular}{rcr}
{\bf (a)}\hskip-2em\raisebox{-\height}{\includegraphics{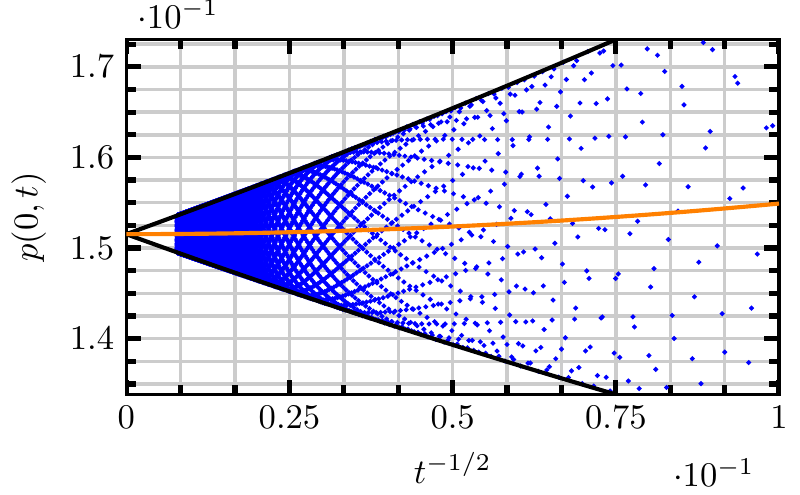}}& ~~ &{\bf (b)}\hskip-2em\raisebox{-\height}{\includegraphics{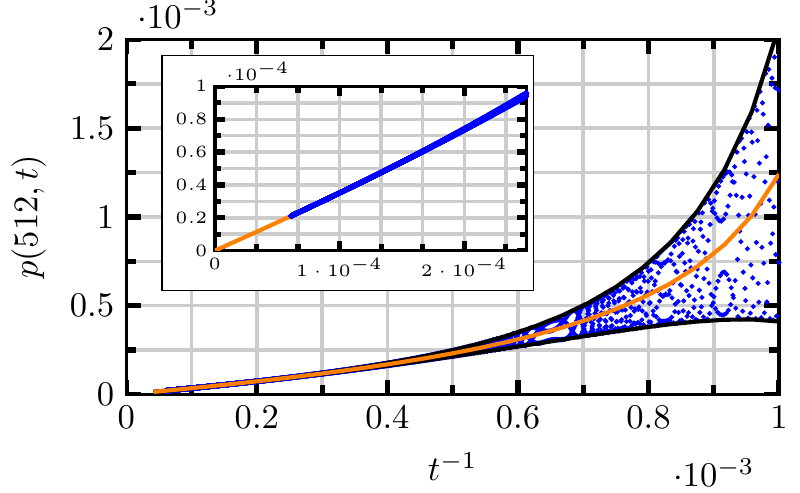}}\\
{\bf (c)}\hskip-2em\raisebox{-\height}{\includegraphics{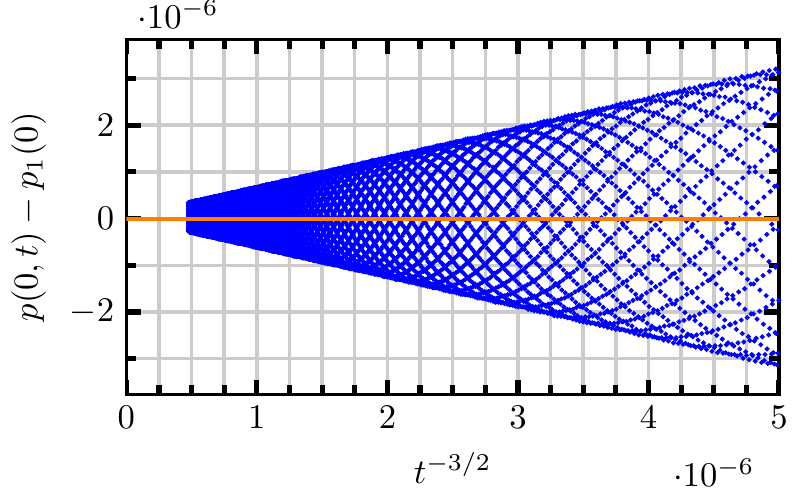}}& ~~ &{\bf (d)}\hskip-1.25em\raisebox{-\height}{\includegraphics{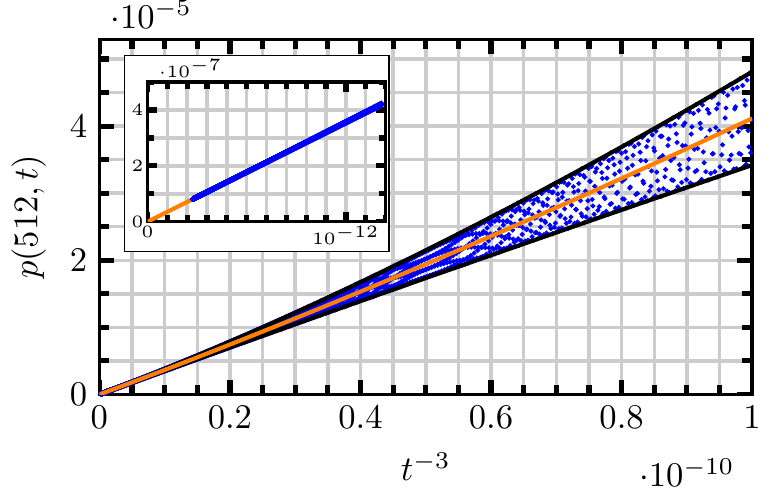}}
\end{tabular}
\caption{The numerically observed probabilities (blue dots), the weak-limit approximation $p_1(n)+p_{avg}$ (orange line), and corresponding envelopes from the long time approximation (black lines). For the upper panels, it is $\ket{\psi_0^0} \propto (0,\I,1)^T$; for the lower ones, it is $\ket{\psi_0^0} \propto (1,0,-1)^T$. The left panels refer to the initial site, whereas the right ones to $n=512$, to contrast  sites with and without significant localization. Note the difference in scaling with time, depending on the initial conditions. The insets depict a zoom for large times on the same $t$-scale. For the data in panel (c), there is no prediction from the method of stationary phase, because for $n=0$ all available orders cancel. However, from the formulas we know that the envelope should scale $\sim t^{-3/2}$, as support by the numerical data.}
\label{fig:comparison_data_pred_p_vs_t_2}
\end{figure*}

\subsection{Convergence}

In the previous sections, we have shown that the 3-state quantum walk on the line is well described by a time independent, localized part, and a ballistically moving front that can be approximated by a smooth PDF. In this section, we investigate, how fast the error of this approximation decays with time.

From the method of stationary phase, we already identified that the localization originates from the $\mat U_1^\dagger\cdot\mat U_1$ term, whereas $p_{avg}$ stems from $\mat U_2^\dagger \cdot \mat U_2 + \mat U_3^\dagger \cdot \mat U_3$. We group the missing terms of the approximation into two functions:
\begin{align}
q_{1} &= \bra{\psi_0^0} \mat U_1^\dagger \cdot (\mat U_2 + \mat U_3) +  (\mat U_2 + \mat U_3)^\dagger \cdot \mat U_1 \ket{\psi_0^0}
\label{eq:p_1_osc}\\
q_{avg} &= \bra{\psi_0^0} \mat U_2^\dagger\cdot \mat U_3 + \mat U_3^\dagger\cdot \mat U_2 \ket{\psi_0^0}
\label{eq:p_avg_osc}
\end{align}
These functions are not non-negative quantities, hence cannot be interpreted as probabilities. In fact, both functions oscillate and average to zero.  As Grimmet {\it et. al.}~\cite{grimmett_2004a} already pointed out, the convergence to the smooth probability function depends in general on the initial conditions. With our approximation, we can determine the slowest convergence rate.

From the formulas in the supplementary material \cite{supplementary_material}, we  see that the leading order of $q_1$ generically is $\sim t^{-1/2}$. By choosing special initial condition, one can cancel this term, and achieve a faster convergence $\sim t^{-3/2}$. This term dominates for small $n$, but is exponentially suppressed for large $n$. In that case $q_{avg} \gg q_1$ and a different convergence rate is possible. In fact for $n\gg 1$, the deviation from the smooth PDF decay at least $\sim t^{-3}$, for particular initial conditions even $\sim t^{-5}$.

Figure~\ref{fig:comparison_data_pred_p_vs_t_2} illustrates our findings. It shows the convergence towards $p_1(0) + p_{avg}(0,t)$ for two different initial conditions at two different sites. The smooth PDF is represented by the orange line. The envelopes (black lines) are derived from Eqs.~(\ref{eq:p_1_osc}-\ref{eq:p_avg_osc}) depending on the site.

Our calculations enable us to make statements about the convergence of the PDF towards the limiting distribution. We have already seen, that the oscillations around the stationary value at the initial site decay $\sim t^{-1/2}$. This was due to the contribution of $\mat U_1(n)$. But for sites sufficiently far away from the initial site, this term becomes exponentially small, and the asymptotic behavior changes. The right panels of  Fig.~\ref{fig:comparison_data_pred_p_vs_t_2} present data similarly to the left, but for $n=512$. By changing the $x$-axis to $t^{-1}$, it is evident from the inset  that $p(512,t) \sim t^{-1}$ for sufficiently large times. This corresponds to the stationary distribution itself. By computing the envelope, we find that the next order correction  vanishes $\sim t^{-3}$, resulting in a correction $\mathcal O(t^{-2})$ for the stationary distribution.

\section{Conclusion}
\label{sec:conclusion}

We studied the 3-state quantum walk on the line in the long time limit using the method of stationary phase. We found explicit formulas for the localization, and found interesting cases where it is only zero for either positive or negative site indices. We showed how the weak limit of the PDF can be interpreted as an time average at a fixed site, or as a spatial average for a fixed $t$. We used the latter interpretation to demonstrate the good agreement between the asymptotic approximation and long time simulations. We applied the smooth, approximative PDF to show that the quantum walk always spreads ballistically for all initial conditions. Finally, we studied the convergence towards this smooth description. We identified the generic convergence rate depending on the site index, and pointed out that other initial conditions only converge faster.

\section{Acknowledgments}
\label{sec:acknowledgments}

We gratefully acknowledge helpful discussions with R. Portugal. This work was supported by DMR-grant \#1207431 from the NSF.
\bibliography{sfalkner-qrw}

\appendix

\section{The stationary probability distribution}
\label{sec_a:localization}
The non-trivial limit of the PDF as $t\to\infty$ is purely determined by the $\kappa$-independent eigenvalue of $\tilde{\mat C}$ in Eq.~(\ref{eq:C_diagonal}). To find this ``stationary state'', we evaluate the definition of $\mat U_1$ in Eq.~(\ref{eq:stationary_inverse_fourier}). For this calculation, we do not have to resort to any approximation, but can solve the integrals analytically by applying the residue theorem.  As already mentioned in the text, all components share the two poles
\[
\kappa_{\pm} = -5 \pm 2 \sqrt{6}
\]
of which only $\kappa_+$ lies inside the unit circle. Depending on $n$, there is an additional pole at $\kappa=0$. After some simple algebra, we find 
\begin{align}
\mat U_{1}(n) &= 
\kappa_{+}^n
\begin{bmatrix}
 \frac{1}{\sqrt{6}} & 1-\sqrt{\frac{2}{3}} & 2-\frac{5}{\sqrt{6}} \\
 -1-\sqrt{\frac{2}{3}} & -\sqrt{\frac{2}{3}} & 1-\sqrt{\frac{2}{3}} \\
 -2-\frac{5}{\sqrt{6}} & -1-\sqrt{\frac{2}{3}} & \frac{1}{\sqrt{6}} \\
\end{bmatrix}
\label{eq:u1_residue}
\\
\notag
& \kern1cm + \Res\limits_{\kappa=0}(\kappa^{n-1}\cdot \tilde{\mat{M}}_1)
\end{align}
Please refer to the supplementary material for the full expression of $\tilde{\mat M}_1$ \cite{supplementary_material}. The last term is only non-zero if $n\leq0$ and counteracts the divergence of $\kappa_+^n$ as $n\to-\infty$. All residues are of the Form
\[
\Res\limits_{\kappa=0}\left( \frac{a \kappa^m}{1+10\kappa+\kappa^2} \right)\,.
\]
To calculate the residue, note that with partial fractions
\allowdisplaybreaks
\begin{align*}
\frac{1}{1+10\kappa+\kappa^2} &= \frac{1}{4\sqrt{6}}\left[ \frac{1}{\kappa_- - \kappa} - \frac{1}{\kappa_+ - \kappa} \right]\\
&= \frac{1}{4\sqrt{6}} \left[ \sum_{k=0}^\infty \frac{\kappa^k}{\kappa_-^{k+1}}
  - \sum_{k=0}^\infty \frac{\kappa^k}{\kappa_+^{k+1}} \right]\,.
\end{align*}
With this representation, we find
\begin{align*}
\Res\limits_{\kappa=0}&\left( \frac{a\kappa^m}{1+10\kappa+\kappa^2} \right) = \frac{a}{2\pi\I} \ointop \kappa^m \frac{1}{1+10\kappa+\kappa^2} \,\D \kappa\\
&=\frac{a}{4\sqrt{6}} \sum_{k=0}^{\infty} \left( \kappa_-^{-k-1} - \kappa_+^{-k-1} \right) \cdot \underbrace{\frac{1}{2\pi\I}\ointop \kappa^{m+k}\D \kappa}_{=\delta_{-1,m+k}}\\
&= \frac{a}{4\sqrt{6}} \left[ \kappa_-^{m} - \kappa_+^m \right]\,.
\end{align*}
Plugging this result into Eq.~(\ref{eq:u1_residue}), yields Eq.~(\ref{eq:u1_final}).

\section{Approximation for long times}
\label{sec_a:asymptotic_approximation}
The inverse Fourier transform reads
\begin{align*}
\ket{\psi_n^t} &= \left(\frac{1}{2\pi}\intop_{-\pi}^{\pi} \EXP{\I \cdot n \cdot k} \cdot \tilde{\mat C}^t\,\D k\right) \cdot \ket{\psi_0^0}\\
 &= \left[\mat U_1(n) + \mat U_2(n,t) + \mat U_3(n,t)\right]\cdot \ket{\psi_0^0}\,,
\end{align*}
We have already seen how $\mat U_1(n)$ emerges from the constant eigenvalue, and how it can be calculated explicitly. For the evaluation of (\ref{eq:def_U23}), we have to resort to an asymptotic analysis for $t\to\infty$.
%
Note that the integrals can be written as
\[
\frac{1}{2\pi}\intop_{-\pi}^{\pi} f(k) \EXP{\I\cdot n \cdot k \pm \I\cdot t\cdot  \tilde\omega} ~\D k
\]
where $f(k)$ represents the components of $\tilde{\mat{M}}_{2,3}$. Before we can apply the method, we introduce the parameter $v$, see Eq.~(\ref{eq:def_v}). This allows us to write the integrals in the form
\[
\frac{1}{2\pi}\intop_{-\pi}^{\pi} f(k) \EXP{\I \cdot t \cdot \left( v \cdot k \pm \omega(k)\right)} ~\D k := \frac{1}{2\pi}\intop_{-\pi}^\pi f(k) \EXP{\I\cdot t \cdot \rho(k)}~\D k
\]
The idea is now to expand $\rho(k)$ around any extrema $k^*$ to second order:
\[
\rho(k) = \rho(k^*) + \frac12 \rho''(k^*) (k-k^*)^2 + \mathcal{O}\left( (k-k^*)^3 \right)\,.
\]
For $t\to\infty$, this captures the main contribution around these points of stationary phase, and everything else is exponentially suppressed. This requires $f(k^*) \neq 0$ which holds for all integrals considered here.  Within this scheme the integral is approximated by
\[
\int f(k) \EXP{\I\cdot t\cdot \rho(k)}~\D k \approx f(k^*) \EXP{\I\cdot t\cdot \rho(k^*)} \int \EXP{\frac{\I \cdot t \cdot \rho''(\kappa^*)}{2} \cdot (k-k^*)^2}~\D k\,,
\]
and the remaining Gaussian integral can be computed exactly. Caution has to be taken with the additional rotation by $\pi/4$ to transform the exponent into the real domain. The direction depends on the sign of $\rho''(k^*)$. This rotation turns the original integration path into a steepest descent on where $\vert \rho'' \vert$ varies the most.

The extrema occur at $k^* = \pm\arccos\left( \frac{1-5v^2}{v^2-1} \right)$ depending on the eigenvalue at hand and the sign of $v$. There is always one such point for each eigenvalue. Furthermore, we find the simple expression:
\[
\rho''(k^*) = \pm \frac{\sqrt{2}}{4}~\sqrt{1-3v^2}~(1-v^2)\,.
\]
The expressions for $\tilde{\mat{M}}_{2,3}(k^*)$ can be found in the Mathematica file \cite{supplementary_material}. They still contain case distinctions for the sign of $n$, which disappears when calculating the expression for $p_{avg}$ in Eq.~(\ref{eq:matrix_for_p_avg}).

\section{Calculating Moments}
\label{seq_a:moment}
Assuming the $p_{avg}(n,t)$ is always a good approximation and that the localized part does not contribute to the time dependence of any moment, we calculate the first three moments of the PDF, $\langle n^k\rangle$ for $k=0,1,2$.  But instead of performing sums over all $n$, we approximate them by integrals 
\begin{align*}
\langle f(n)\rangle &= \sum_{n=-\infty}^{\infty} f(n) p(n,t) \\
&\approx \intop_{-1/\sqrt{3}}^{1/\sqrt{3}} f(v\cdot t) p_{avg}(v\cdot t, t) \cdot t ~ \D v
\end{align*}
Applying this to every matrix entry in Eq.~(\ref{eq:matrix_for_p_avg}) yields:
\begin{align}
&\langle n^0\rangle = \frac{1}{\sqrt{6}}~\bra{\psi_0^0}
\begin{bmatrix}
-1+\sqrt{6} & 2-\sqrt{6} & 5-2\sqrt{6}\\
2-\sqrt{6}  &      2     & 2-\sqrt{6}\\
5-2\sqrt{6} & 2-\sqrt{6} & -1+\sqrt{6}
\end{bmatrix} 
\ket{\psi_0^0}
\label{eq:zeros_moment}\\
&\begin{aligned}
\langle n\rangle &= \frac{t}{\sqrt{6}}\\
&\bra{\psi_0^0}
\begin{bmatrix}
 2-\sqrt{6} & -2+\sqrt{6} &     0     \\
-2+\sqrt{6} &      0      & 2-\sqrt{6}\\
    0       & 2-\sqrt{6} & -2+\sqrt{6}
\end{bmatrix} 
\ket{\psi_0^0}\end{aligned}
\label{eq:first_moment}\\
&\begin{aligned}
\langle n^2\rangle &=\frac{t^2}{6\sqrt{6}}\\
&\bra{\psi_0^0}\begin{bmatrix}
-13+ 6\sqrt{6} & 14-6\sqrt{6} & 29-12\sqrt{6} \\
 14- 6\sqrt{6} &       2      & 14-6\sqrt{6}  \\
 29-12\sqrt{6} & 14-6\sqrt{6} &-13+6\sqrt{6}
\end{bmatrix} 
\ket{\psi_0^0}
\end{aligned}
\label{eq:second_moment} 
\end{align}
We observe that the zeros moment is unity only  for the initial conditions that show no localization, $\bra{\psi_0^0}\propto(1,-2,1)^T$. The matrix for the first moment has  the eigenvector $(1,1,1)^T$ with eigenvalue $0$. However, this does not cover all symmetric initial conditions that will yield a zero first moment by symmetry. It easily verified that the initial condition $\sim (1,0,1)^T$ also yield a zero first moment. Hence, every linear combination of those two will do so, too, which now covers all symmetric initial conditions. These asymptotic formulas show that any non-zero first moment  grows linearly in time  while the second moment is proportional to $t^2$.
\end{document}